\pdfoutput=1
\documentclass[twocolumn,superscriptaddress,aps,preprintnumbers,amsmath,amssymb,pral]{revtex4-1}
\usepackage{graphicx}
\usepackage{bm}
\usepackage{color}
\usepackage{amsmath,amssymb}	
\usepackage{hyperref}
\usepackage{setspace}
\usepackage{longtable}
\usepackage{booktabs}
\usepackage{comment}
\usepackage{array}
\usepackage{cancel}
\usepackage{enumitem}
\def\slashchar#1{\setbox0=\hbox{$#1$} % set a box for #1
\dimen0=\wd0 % and get its size
\setbox1=\hbox{/} \dimen1=\wd1 % get size of /
\ifdim\dimen0>\dimen1 % #1 is bigger
\rlap{\hbox to \dimen0{\hfil/\hfil}} % so center / in box
#1 % and print #1
\else % / is bigger
\rlap{\hbox to \dimen1{\hfil$#1$\hfil}} % so center #1
/ % and print /
\fi}

\def\beq{\begin{eqnarray}}
\def\eeq{\end{eqnarray}}

\newcommand{\lsim}{ \mathop{}_{\textstyle \sim}^{\textstyle <} }

\newcommand{\MEV}{ {\rm MeV} }
\newcommand{\GEV}{ {\rm GeV} }
\newcommand{\TEV}{ {\rm TeV} }
\newcommand{\CM}{ {\rm cm} }
\newcommand{\MM}{ {\rm mm} }

\newcommand{\FBI}{ {\rm fb}^{-1} }
\newcommand{\ABI}{ {\rm ab}^{-1} }

\newcommand*{\pt}{p_{\rm T}}
\newcommand*{\MET}{E_{\rm T}^{\rm\,miss}}

\def\SU{\mathop{\rm SU}}

\newcommand{\MOT}[1]{#1}

\begin{document}
\newcolumntype{Y}{>{\centering\arraybackslash}p{23pt}} 

%%%%%%%%%%%%%%%%%%%%%%%%%%%%%%%%%%
%%%%%%%%%%% Title page %%%%%%%%%%%
%%%%%%%%%%%%%%%%%%%%%%%%%%%%%%%%%%

\preprint{KYUSHU-RCAPP-2019-05, UT-19-24, IPMU19-0145}

\title{Cornering Higgsino: Use of Soft Displaced Track 
}

\author{Hajime Fukuda}
\affiliation{Theoretical Physics Group, Lawrence Berkeley National Laboratory,
 CA 94720, USA}
\affiliation{Berkeley Center for Theoretical Physics, Department of Physics,\\
 University of California, Berkeley, CA 94720, USA}

\author{Natsumi Nagata}
\affiliation{Department of Physics, University of Tokyo, 
Tokyo 113-0033,
Japan}
%\email{natsumi@hep-th.phys.s.u-tokyo.ac.jp} 
 
 \author{Hideyuki Oide} 
\affiliation{Department of Physics, Tokyo Institute of Technology, Tokyo 152-8551}
 
\author{Hidetoshi Otono} 
\affiliation{Research Center for Advanced Particle Physics, Kyushu University, Fukuoka 819-0395, Japan}
%\email{Hidetoshi.Otono@cern.ch}

\author{Satoshi Shirai}
%\email[e-mail: ]{satoshi.shirai@ipmu.jp}
\affiliation{Kavli IPMU (WPI), UTIAS, The University of Tokyo, Kashiwa, Chiba 277-8583, Japan}

\date{\today}
\begin{abstract}
Higgsino has been intensively searched for in the LHC experiments in recent years. Currently, there is an uncharted region beyond the LEP Higgsino mass limit where the mass splitting between the neutral and charged Higgsinos is around $0.3\hbox{--}1~\GEV$, which is unexplored by either the soft di-lepton or disappearing track searches. This region is, however, of great importance from a phenomenological point of view, as many supersymmetric models predict such a mass spectrum. In this letter, we propose a possibility of filling this gap by using a soft micro-displaced track in addition to the mono-jet event selection, which allows us to discriminate a signature of the charged Higgsino decay from the Standard Model background. It is found that this new strategy is potentially sensitive to a Higgsino mass of $\lesssim 180$ (250)~GeV at the LHC Run2 (HL-LHC) for a charged-neutral mass splitting of $\simeq 0.5$~GeV.

\end{abstract}

\maketitle

\section{Introduction}
\label{sec:introduction}

The absence of a Dark Matter (DM) candidate in the Standard Model (SM) invokes the strongest motivation for the extension of the SM.
Among various DM candidates, Weakly Interacting Massive Particles (WIMPs) are the most promising and well-motivated, since the correct DM abundance can be obtained via the conventional freeze-out mechanism~\cite{Bernstein:1985th,Srednicki:1988ce} and many popular extensions of the SM provide such a particle.

Higgsino---the fermionic partner of the Higgs fields in the supersymmetric SM (SSM)---is a well-known example for a WIMP. \MOT{If the Higgsino is the lightest supersymmetric particle and its mass is $\lesssim 1$~TeV, its thermal relic abundance does not over-close  the Universe \cite{Cirelli:2007xd},} making it a promising candidate for WIMP DM. A similar DM candidate, a stable SU(2)$_L$ doublet fermion, is also frequently considered in the bottom-up approach;
systematic classification of DM candidates in terms of gauge charges reveals that the doublet fermion is the simplest possibility for gauge-portal DM~\cite{Cirelli:2005uq,Cirelli:2007xd,Cirelli:2009uv}. Such a DM candidate is, therefore, a crucial target for the DM hunting. In this context those are collectively referred to as ``Higgsino'' in what follows.

In SSMs, the Higgsino mass is connected to the scale of the electroweak symmetry breaking, and thus the \emph{naturalness} argument requires it to be around the weak scale~\cite{Chan:1997bi}. Moreover, in some cases, the Higgsino mass is radiatively generated from the gaugino loops and suppressed by a loop factor compared to the gaugino masses~\cite{Hall:2011jd}. These theoretical considerations suggest that Higgsino is rather light and thus potentially accessible by the LHC experiments.

There are various alternative ways to detect the Higgsino DM. For example, 
the DM direct/indirect detection experiments have already put some constraints on the Higgsino DM \MOT{(see, e.g., Ref.~\cite{Kowalska:2018toh})}.
These experiments, however, suffer from large astrophysical uncertainties~\cite{Ullio:2016kvy,Benito:2016kyp,Ichikawa:2016nbi,*Ichikawa:2017rph,Calore:2018sdx}. Moreover, these constraints are usually obtained on the assumption that DM is solely composed of Higgsino, and would be significantly weaker if this is not the case. Collider experiments are, on the other hand, free from such uncertainties, and thus DM searches at colliders can offer the most conservative test for the Higgsino DM scenario.

The Higgsino search strategy at colliders strongly depends on the mass splitting among the Higgsino DM and its isospin partners.
If the mass splitting is greater than $O(1)~\GEV$, Higgsinos can be probed with multi-lepton signals~\cite{Gori:2013ala,Han:2014kaa,Bramante:2014dza,ATLAS:2019lov}. For small mass difference $\lesssim 300~\MEV$, on the other hand, the charged Higgsino can be long-lived and detected \MOT{in the disappearing charged track searches~\cite{Low:2014cba,ATL-PHYS-PUB-2017-019}. }
Improvement of the tracking techniques at future colliders may enable to probe even $1~\TEV$ Higgsinos with this strategy \cite{Mahbubani:2017gjh, Fukuda:2017jmk}.

Meanwhile, the intermediate region, i.e., mass splitting of around $0.3-1~\GEV$, has never been probed at the LHC. 
This letter proposes a new strategy to explore this region by utilizing a soft pion from the decay of a charged Higgsino \footnote{
The soft track was extensively used in the DM search at the LEP experiments~\cite{LEP:Higgsino}.
The use of the soft tracks are also discussed for future lepton-lepton colliders~\cite{Berggren:2013vfa} and lepton-hadron colliders~\cite{Curtin:2017bxr}.
},
which has a barely discernible lifetime with the LHC tracker resolution, and demonstrates such a micro-displaced pion can be an effective signature to distinguish the signal events from the SM background.

\section{Higgsino phenomenology}
Higgsino $\tilde H_{u,d}$ is a pair of $\SU(2)_L$ doublet Weyl fermions with hypercharge $Y = \pm 1/2$.
Conventionally, we take the mass term  ${\cal L} \ni - \mu  \tilde{H}_{u} \tilde{H}_{d} + {\rm h.c.}$ with real and positive $\mu$  without loss of generality.

The completely pure Higgsino DM has already been excluded, as the DM is a Dirac fermion and the $Z$-boson-mediated nucleon scattering cross section is far above the current direct detection limits.
For the Higgsino DM to be viable, therefore, some new particles heavier than the Higgsino are required. In the minimal SSM (MSSM), the mixing with gauginos splits the Higgsino into Majorana fermions, for which the direct detection limits are avoided~\cite{Nagata:2014wma,Nagata:2014aoa}.

The Higgsino doublet is decomposed into two neutral Majorana fermions ($\chi^0_1, \chi^0_2$) and a charged fermion ($\chi^{\pm}$). We adopt the convention $\Delta m_0 \equiv  m_{\chi^0_2} - m_{\chi^0_1}>0$, and  define
 $\Delta m_\pm \equiv  m_{\chi^\pm} - m_{\chi^0_1}$.
$\Delta m_0$ is mainly induced via the tree-level mixing with heavier particles---bino or wino in the MSSM---and can be approximated by
\begin{align}
   \Delta m_0^\text{tree} &\simeq  M_Z^2\left|\frac{c_W^2}{M_2} + \frac{s_W^2}{M_1}\right|,
\end{align}
where $c_W(s_W)=\cos \theta_W (\sin \theta_W)$ with $\theta_W$ the Weinberg angle, and $M_Z, M_1, M_2$ are the masses of the $Z$ boson, bino, and wino, respectively. 
The charged-neutral mass difference receives electroweak radiative corrections as well, being expressed as
$\Delta m_{\pm} = \Delta m_{\pm}^{\rm rad} +  \Delta m_{\pm}^{\rm tree}$ with 
\begin{align}
    \Delta m_\pm^\text{rad} &\simeq \frac{\alpha_2 s_W^2 \mu}{2\pi}  \int^1_0  {\rm d}t(1+t)\ln
\left[1+\frac{M_Z^2(1-t)}{\mu^2 t^2}\right], \\
    \Delta m_\pm^\text{tree} &\simeq \frac{\Delta m_0^\text{tree}}{2}    + \sin2\beta M_Z^2\left(\frac{c_W^2}{M_2} - \frac{s_W^2}{M_1}\right),
\end{align}
where $\alpha_2$ is the fine structure constant for the $\text{SU}(2)_L$ gauge interaction and  $\tan\beta \equiv \langle  H_u \rangle /\langle H_d \rangle$. For $\mu \gg M_Z$, 
\begin{align}
    \Delta m^\text{rad}_\pm \simeq \frac{\alpha_2 M_Z}{2}\sin^2\theta_W \simeq 354\,\text{MeV}.
\end{align}
Note that the above estimations can be modified with radiative corrections  by $O(10)$\%  \,\cite{Nagata:2014wma}.

In the MSSM, the mass splittings get larger for a larger Higgsino-gauginos mixing.
This mixing also increases the DM-nucleon scattering cross sections and electronic dipole moments (EDMs) of the SM fermions. As a result, there is a strong correlation between the Higgsino mass splittings and these observables \cite{Fukuda:2017jmk}, and in particular, large mass splitting regions are testable in DM direct detection and EDM experiments. 

To see this correlation, let us consider the Higgsino-nucleon elastic scattering, which is dominantly induced by the tree-level Higgs-boson exchange process. If, e.g., the gaugino masses are real and obey the so-called GUT relation, $M_{1}/M_2=\alpha_1/\alpha_2$, we find a simple relation between the charged-neutral mass splitting $\Delta m_{\pm}$ and the spin-independent DM-proton scattering cross section $\sigma^{\rm SI}$:
\begin{align}
    \Delta m_{\pm} \sim
     \Delta m^\text{rad}_{\pm} + 170~{\rm MeV} \left( \frac{\sigma^{\rm SI}}{10^{-48}~{\rm cm}^2} \right)^{1/2},
\end{align}
for $\tan\beta \gg 1$.
The current experimental bound on $\sigma^{\rm SI}$ is $\sim 10^{-46}~ {\rm cm}^2 ( m_{\chi^0_1}/100~\GEV)$  for the DM local density  $\rho_{\chi^0_1} = 0.3~{\GEV/\CM^3}$~\cite{Aprile:2018dbl}.
The present limit has already imposed a limit $\Delta m_{\pm}^{\text{tree}}>O(1)~\GEV$. Nevertheless, the region of  $\Delta m^{\rm tree}_\pm <O(100)~\MEV$ cannot be probed in direct detection experiments even if their sensitivities are improved down to the neutrino floor.

A similar argument can also be made for EDMs.
If the gaugino-Higgsino system has CP violation, EDMs are induced at two-loop level \cite{Barr:1990vd, Chang:2005ac, *Deshpande:2005gi, *Giudice:2005rz}.
The size of the EDMs and the mass splitting $\Delta m_\pm^{\rm tree}$ are also correlated.
It is, however, difficult to probe the region $\Delta m_\pm^{\rm tree}<O(100)~\MEV$ even with future experiments \cite{Fukuda:2017jmk}.

Therefore, it is important to uncover the parameter region $\Delta m_\pm^{\rm tree}<O(100)~\MEV$ at colliders.
The Higgsino phenomenology at  colliders is sensitive to the decay of the heavier Higgsino components,
which significantly depend on the mass splittings.
If, in particular, $m_{\pi^{\pm}} < \Delta m_{\pm} \ll 1$~GeV, with $m_{\pi^\pm}$ the mass of the charged pion $\pi^\pm$, the main decay mode of the charged Higgsino is  ${\chi}^\pm \to \chi^0_{1,2} \pi^{\pm}$ \cite{Feng:1999fu}. The partial decay length of ${\chi}^\pm \to \chi^0_{1} \pi^{\pm}$ is approximately given by \cite{Chen:1995yu,*Chen:1996ap,*Chen:1999yf, Thomas:1998wy} 
\begin{align}
\label{eq:ctau}
\Gamma^{-1}_{{\chi}^\pm \to \chi^0_{1} \pi^{\pm}  } \simeq \frac{14~\text{mm}}{\hbar c}\times \left[\left(\frac{\Delta
					m_\pm}{340\,\text{MeV}}\right)^3\sqrt{1
- \frac{m_{\pi^\pm}^2}{\Delta m_\pm^2}}\right]^{-1}.
\end{align}
The decay ${\chi}^\pm \to \chi^0_{2} \pi^{\pm}$ is also possible if $\Delta m_{\pm} - \Delta m_0 > m_{\pi^\pm}$ and this decay rate can be obtained by replacing $\Delta m_{\pm}$ by $\Delta m_{\pm} - \Delta m_0$ in Eq.~\eqref{eq:ctau}.
\MOT{For larger $\Delta m_\pm$, three-body decay modes open up, in which 
one charged particle, such as a charged lepton or meson, is emitted.
We also include these decay modes in the following analysis.}

If the tree-level mass difference is small enough, $\Delta m_\pm^{\rm tree} \ll 100~\MEV$, the decay length of the charged Higgsino is sufficiently long, $O(1)~\CM$, so that the disappearing charged track search can test the Higgsino DM \cite{Fukuda:2017jmk}, as in the case of the wino DM \cite{Ibe:2006de, Buckley:2009kv, Asai:2007sw, *Asai:2008sk, *Asai:2008im,Saito:2019rtg}.

\section{The soft-displaced \MOT{track} signal from the Higgsino decay at LHC}
Higgsinos can be pair-produced via the Drell-Yan process at the LHC, which eventually yield a pair of the lightest Higgsinos $\chi_{1}^{0}$, the DM candidate.
The  ``mono-jet search'' is regarded as a model-independent search for DM production at the LHC, where the signal is an invisible pair-production of DM particles accompanied by a jet from initial-state radiation (ISR). 
This search is, however, not sensitive to Higgsinos \cite{Han:2013usa,Ismail:2016zby}; as illustrated in the re-interpretation of the ATLAS mono-jet search performed in Ref.~\cite{Aaboud:2017phn}, currently it doesn't give any constraints on the Higgsino DM.

Adding distinctive signatures improves the sensitivity in specific parameter spaces. For instance, if  $\Delta m_0 > O(1)~\GEV$, soft di-leptons from the heavier neutral Higgsino $(\chi_{2}^{0})$ decays become an important discriminant of the Higgsino signals. Meanwhile, if $\Delta m_{\pm}$ is smaller than $\sim0.3~\GEV$, the charged Higgsino $(\chi^{\pm})$ can be long-lived, as shown in Eq.~(\ref{eq:ctau}), and it may be detected as a disappearing charged track. Both search strategies have been used in ATLAS~\cite{PhysRevD.97.052010,Aaboud:2017mpt} and CMS~\cite{Sirunyan2018,*Sirunyan2018-disapp}.

Remarkably, the parameter region $0.3\lsim \Delta m_{\pm} \lsim 1~\GEV$ has never been explored at the LHC, and the LEP still gives the strongest constraint \cite{LEP:Higgsino}. 
This is due to a lack of distinctive signatures to be added to the mono-jet event topology in this region.
We, however, notice that compared to the ``disappearing track'' regime, \MOT{the charged meson and lepton from the $\chi^{\pm}$ decay} in this region can be hard enough to surpass the track reconstruction momentum threshold of $500~\MEV$, while the charged Higgsino lifetime still remains discernible by the track's displacement from the primary $pp$ interaction. As we see below, this signature can distinguish the Higgsino events from the SM background and thus offer a promising way of filling ``$\Delta m_{\pm}$ gap'' at the LHC.

The performance of the ATLAS detector~\footnote{Following the standard hadron-collider convention, we take the cylindrical coordinate system $(r,z,\phi)$, where $\phi$ is the azimuthal angle in the plane transverse to the beam direction.
The track parameters are defined as follows:
$\pt$: transverse momentum;  $\theta$: polar angle; the pseudorapidity $\eta = -\ln\tan(\theta/2)$; $d_{0}$:  transverse impact parameter with respect to the beam centroid; $\Delta z_{0}$: longitudinal impact parameter with respect to a primary vertex position.} is semi-quantitatively mimicked as follows: the number of pileup vertices is assumed to be $\langle \mu  \rangle = 35~ {\rm and}~  200$ for Run2 and  high-luminosity LHC (HL-LHC)  respectively;
\MOT{
tracks are assumed to be reconstructed with the standard tight track selection and its reconstruction efficiency for charged particles is conservatively assumed to be 80\% for $\pt>500$ MeV and $|\eta|<2.5$; charged particles not satisfying these conditions are discarded;
the fake-track rate in the standard tight track selection is negligibly small \cite{ATL-PHYS-PUB-2015-051, ATLAS:FAKE};} the resolution of the transverse impact parameter is parameterized as
$
    \sigma_{d_{0}}(\pt)~[{\rm mm}] =0.01 +  0.08/(\pt/\GEV),
$
and we set $\sigma_{\Delta z_{0}} = 5\,\sigma_{d_{0}}$ for Run2 and $\sigma_{\Delta z_{0}} =  \sigma_{d_{0}}$ for HL-LHC. \cite{ATLAS:IP}. This configuration takes account of the installation of a factor-5 finer resolution pixel detector in $z$-direction for HL-LHC in ATLAS~\cite{ATLAS-TDR-030}. It is important to take into account 
the non-Gaussianity of the impact parameter resolution due to multiple Coulomb scattering and secondary particles by nuclear interaction with the detector material. To incorporate this, we use a Crystal Ball function with a Gaussian core up to $\pm 2\sigma$ and power-law tail of slope three.
This treatment can well reproduce the measurement of the ATLAS impact parameter distribution for minimum-bias events inclusive of primary and secondary charged particles~\cite{Aad:2016mok}.

For Monte-Carlo simulations, we use {\sc Madgraph5} \cite{Alwall:2014hca} event generator interfaced to {\sc Pythia8} \cite{Sjostrand:2014zea} parton shower and hadronization, with the detector response simulated by {\sc Delphes3} \cite{deFavereau:2013fsa}. We refer to the NLO-NLL cross sections \cite{Fuks:2012qx,Fuks:2013vua,Beenakker:1996ed}. All of the main SM background contributions in the mono-jet searches are included, and the event yield is adjusted to match with the mono-jet data of Ref.~\cite{Aaboud:2017phn}.

In order to demonstrate the sensitivity gain with respect to the mono-jet search, we take the same event selection criteria as the ATLAS mono-jet search as the baseline selection~\cite{Aaboud:2017phn}: a leading jet with $\pt>250~\GEV$ and $|\eta|<2.4$; up to four jets with $\pt>30~\GEV$ and $|\eta|<2.8$; separation with missing transverse momentum $\Delta \phi({\rm jet},\vec{p}^{\rm\,miss}_{\rm T})>0.4$; leptons are vetoed. The magnitude of the missing transverse momentum, $\MET$, for the signal region is required to be $\MET>500 (700)~\GEV$ for the Run2 (HL-LHC).
With this selection, the main backgrounds are $Z(\to \nu \bar{\nu})$ and $W(\to \ell \nu)$.

\begin{figure}[t]
{\includegraphics[width=0.5\textwidth]{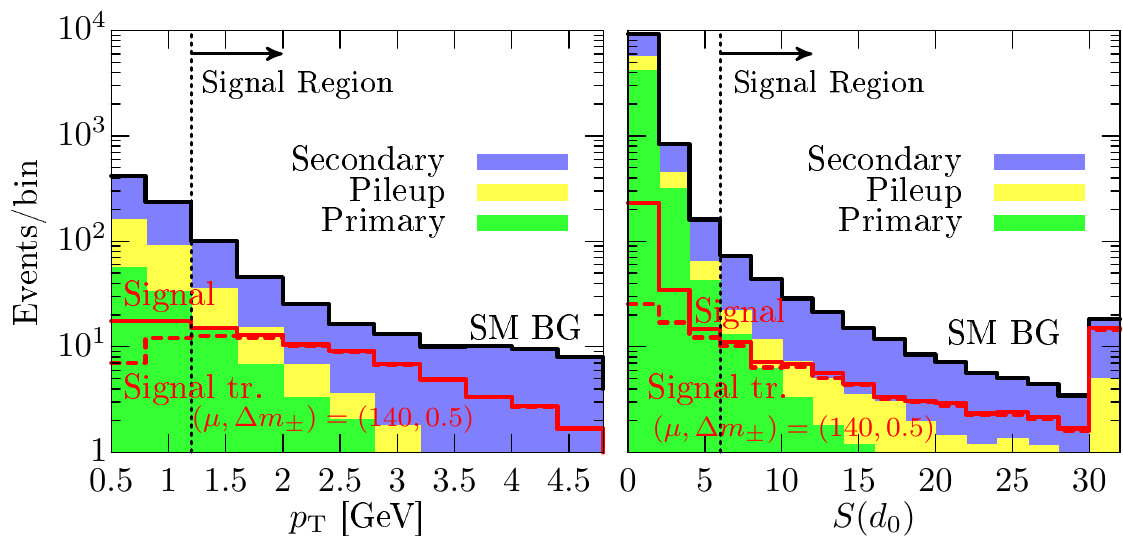}} 
\caption{The signal and SM background distributions in $\pt$ (left) and $S(d_{0})$ (right) after 
the rest of the selection criteria (see the text) are applied for each case, 
for an integrated luminosity of $140~\FBI$.
\MOT{The contributions from primary, secondary  and pileup vertexes are shown in green, blue and yellow, respectively. }
For the signal Higgsino, the case of $m_{\chi^0_1}=140~\GEV$, $\Delta m_{\pm}=500~\MEV$ and $\Delta m_0 = 0~\MEV$ is shown.
The dashed lines show the distribution of the tracks from the true Higgsino decay; the difference between the red solid and dashed lines is due to the SM BG contributions in the signal events. 
}
\label{fig:result_dist}
\end{figure}

In addition to the mono-jet selection, we require at least one extra track satisfying the following conditions, which is expected to be the charged pion from the charged Higgsino decay:
\begin{itemize}
    \item Basic Selection: $1.2~\GEV<\pt<5~\GEV$; $|\eta|<1.5$; $|\Delta z_{0} \sin(\theta)|<1.5~\MM$ and $|d_{0}|<10~\MM$. The candidate must have a hit at the innermost pixel layer, located at $r=33~\MM$ in the ATLAS detector;
    \item Isolation: The candidate track is separated by  $\Delta R \equiv \sqrt{(\Delta \phi)^2 + (\Delta \eta)^2}>1$ for any tracks with $\pt>1~\GEV, ~\Delta z_{0} \sin(\theta)|<1.5~\MM$ and $|d_{0}|<1.5~\MM$;
    \item Displacement: The transverse impact parameter of the candidate is large: $S(d_{0})\equiv |d_{0}|/\sigma_{d_{0}}>6$. This slight displacement arises from a sizable lifetime of the charged Higgsino;
    \item Alignment to $\MET$ direction: $\Delta \phi({\rm trk.},\vec{p}^{\rm\,miss}_{\rm T})<1$.
\end{itemize}

FIG.~\ref{fig:result_dist} shows the signal and background distributions in $\pt$ and $S(d_{0})$ 
after the rest of the selection criteria are applied for each case. 
The sensitivity further improves by, e.g., introducing a multivariate analysis, but such optimization is beyond the scope of this letter. Around 10--20 tracks satisfy the Basic Selection for each background event. In order to quantify the selection and rejection efficiency, we define the track's relative \emph{pass rate} for each selection step as the rate of the tracks which have survived the previous step in the same order as listed above passing this step. 
The pass rates of QCD-origin tracks for the isolation, displacement, and $\MET$-alignment selection steps are approximately 4\%, 2\%, and 40\%, respectively.
The isolation requirement efficiently suppresses tracks from heavy flavor hadrons.
It is observed that tracks from $\tau$ decay tend to pass these selections with a total pass rate of around $1$\%, which is much larger than that for QCD tracks. 
The requirement $\pt < 5~\GEV$ plays a significant role in rejecting the $\tau$ decay products.

For the above track selection, the signal selection efficiency is larger for events with larger $\MET$, as Higgsinos tend to be more boosted by the ISR recoil. 
In the case that $\Delta m_{\pm} \simeq 500$ MeV, the acceptance rate of the signal track requirement is around 5\%.
For the HL-LHC, it reduces to 3\% due to the failure of the isolation selection by larger pileup.
The background event yield passing all the selections is estimated to be around 0.5\% of the events that have passed the mono-jet selection.

\begin{figure}[t]
{\includegraphics[width=0.47\textwidth]{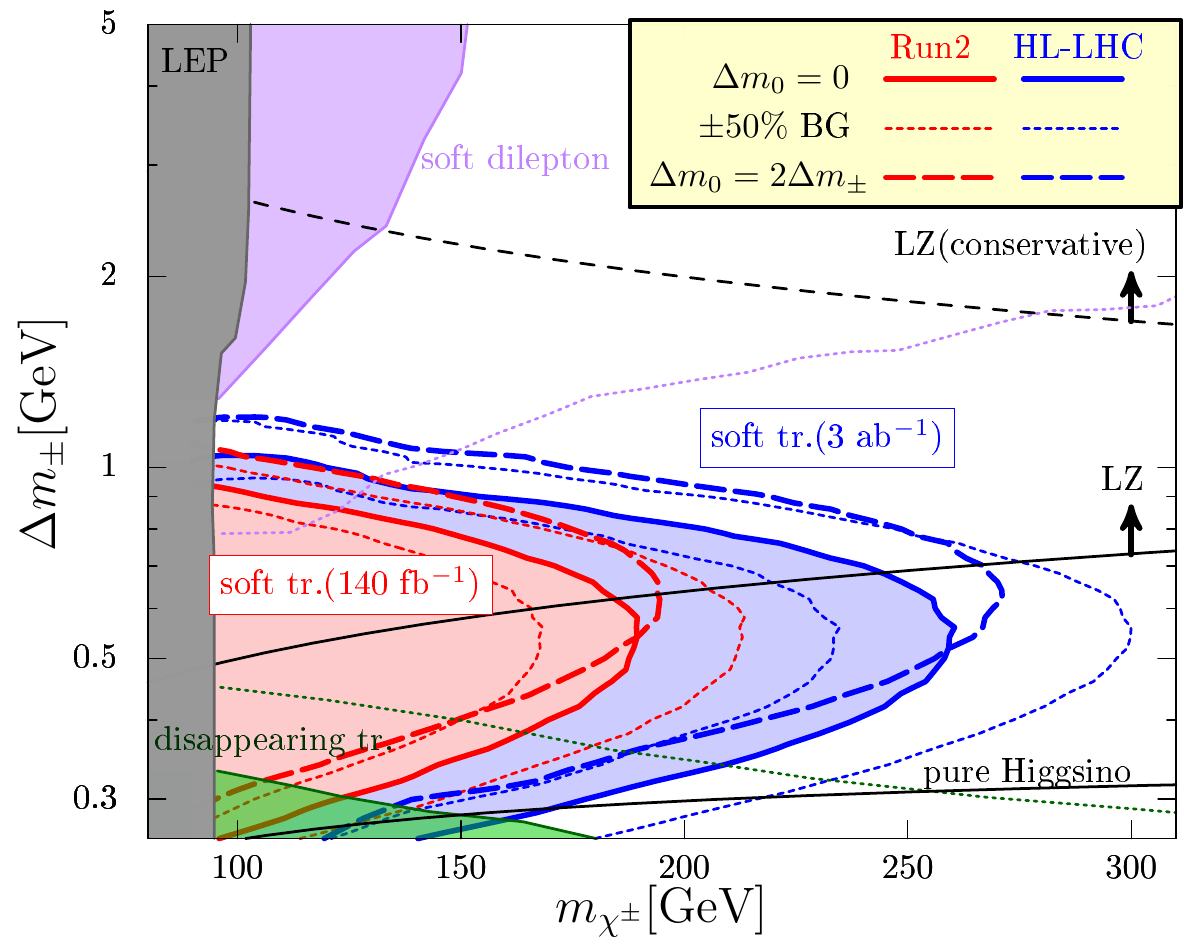}} 
\caption{
The expected reaches of the Higgsino search with our method at the LHC Run2 $140~\FBI$ and HL-LHC  $3~\ABI$ shown in the red and blue areas, respectively, for $\Delta m_0 = 0~\GEV$ (solid) and $2 \Delta m_{\pm}$ (dashed).
The dotted lines show the $\pm50\%$ uncertainty of the background estimation for $\Delta m_0 = 0~\GEV$.
The gray, green, and purple regions are excluded by the LEP \cite{LEP:Higgsino}, the disappearing track search \cite{ATL-PHYS-PUB-2017-019}, and the soft dilepton search \cite{ATLAS:2019lov}, respectively.
\MOT{
The purple and green dotted lines show the HL-LHC prospects of the dilepton and disappearing track searches \cite{ATL-PHYS-PUB-2018-031}, respectively.}
The LZ sensitivity \cite{Akerib:2018lyp} is also shown in the black solid and dashed lines for the cases discussed in the text. 
}
\label{fig:result}
\end{figure}

With this selection, the number of background events is around $250(1000)$ for Run2 $140~\FBI$(HL-LHC $3~\ABI$). Given that the background tracks are largely originated from relatively soft QCD processes as well as from secondary interactions, yield estimation by MC simulation might not be sufficiently accurate, and thus a data-driven background estimation would be more reliable. As a mimic of data-driven background estimation, the ``ABCD method''~\footnote{This method maps the background distribution in two selection variables $x,y$ after all the event selection criteria except these variables are imposed. The distribution can be split into four mutually exclusive regions $\{A,B,C,D\}$ by the event selection criteria on these variables where the regions $(A,D)$ and $(B,C)$ are diagonal respectively, and the signal region selection is one of the four regions (say, the region $A$). The mapping variables are carefully chosen so that the signal contamination in the other three regions is considered to be negligibly small. If the background distribution $f(x,y)$ is orthogonal in these variables, i.e.~the distribution can be modelled as $f(x,y)=g(x)h(y)$ where $g,h$ are arbitrary distributions, then the yield of the signal region $A$ can be estimated using the yields in the other three ``control regions'' by making use of the orthogonality: $N_{A}=N_{B}N_{C}/N_{D}$.} using $\MET$ and $S(d_{0})$ as the two orthogonal variables is attempted. 
\MOT{
For instance, the control and signal regions can be defined with $\MET \lessgtr 500~\GEV$ and $S(d_0)  \lessgtr 6$ at Run2.}
It is found that the statistics of the background events in the control regions are abundant ($>2000$ for each control region), that the orthogonality between these two variables is very good, and that a good closure within statistical uncertainty of around 3\% is obtained.
\MOT{
The remnant backgrounds are secondary particles (decays from $K_S$, strange baryons, etc.) ($\sim 60\%$),
mis-measurement of primary particles ($\sim 20\%$), and pileup ($\sim 20\%$) for the $Z(\to \nu \bar{\nu})$ event at Run2.
In the HL-LHC with $\langle \mu \rangle = 200 $, the pileup contributions increase up to around 60\% and become dominant background.}
For the $W(\to \ell \nu)$ background, the dominant ($\sim 50\%$) contribution arises from $\tau$ decay products.
It is found that this result does not strongly depend on the choice of the {\sc Pythia8} tune. It is also confirmed that the  converted photon contribution\,\cite{SMU,Aaboud:2017pjd,Fukuda:2016qah} is negligible.

In FIG.~\ref{fig:result}, we show the expected reaches of the Higgsino search at the LHC Run2 $140~\FBI$ (red region) and HL-LHC  $3~\ABI$ (blue region).
We adopt the $CL_s$ prescription~\cite{Read_2002} to derive the 95\%$CL_s$ limit, assuming the systematic uncertainty of the background estimation to be 3\%, \MOT{which we infer from our analysis with the ABCD method as discussed in the previous paragraph}.
The solid (dashed) lines show the case of $\Delta m_0 = 0 (2m_{\pm})$.
In the latter case, the lifetime of the charged Higgsino is twice that in the former due to the absence of the decay $\chi^\pm \to \chi_2^0$, and hence a larger mass splitting region can be probed.
For reference, in the dotted lines, we show the variation in the limit when scaling the background yield by $\pm50\%$ for $\Delta m_0 = 0$. 
The current collider constraints are also overlaid \cite{LEP:Higgsino,ATL-PHYS-PUB-2017-019,ATLAS:2019lov}.
It is possible to probe the Higgsino mass up to $180(250)~\GEV$ at Run2 (HL-LHC) for $\Delta m_{\pm}=500~\MEV$ .

We also show in Fig.~\ref{fig:result} the prospects of the future DM direct searches, for which we consider the LZ experiment as an example \cite{Akerib:2018lyp}, where we assume the GUT relation for gaugino masses and $\tan\beta \gg 1$. The black solid line corresponds to the case that the whole DM consists of Higgsinos, while the black dashed line is for a more conservative case that the amount of Higgsino is equal to its thermal relic abundance (and thus the sub-component of the DM
%only a sub-component of the DM density is occupied by Higgsinos
). As we see, the Higgsino search with our method is complementary to the DM direct detection experiments, and thus plays a crucial role in testing the light Higgsino scenario.

\section{Conclusion and Discussion}
In this work, we have explored the possibility of making use of a soft displaced track as a new probe of the Higgsino search at the LHC. It was found that requiring such a track in the ISR-recoiled events with a significant $\MET$ would enable a clean separation of the Higgsino signal from SM backgrounds.
This method allows us to access the parameter region to which the existing LHC searches have never been sensitive beyond the LEP limit. Given that reconstruction of such soft displaced tracks is feasible with the mere use of the established standard tracking method and that the dataset has already been recorded, the installation of this search method to the LHC experiments is to be an urgent task. 

Tagging soft displaced tracks is useful for not only Higgsino but also other DM models.
For instance, in the sfermion-neutralino coannihilation region, the sfermion decay emits such soft particles.
Another example is the 5-plet minimal DM case \cite{Cirelli:2005uq}, for which 
the decay length of the doubly charged particle is $\sim 1~\MM$ and thus the present method is applicable.

In the present work, we have focused on Higgsinos with a decay length of $\lesssim10~\MM$. 
For a longer decay length, we can consider combination of a disappearing charged track and a soft track as a ``kink'' signature, which may extend the sensitivity further. Technical development in this direction is indeed present within ATLAS~\cite{ATL-PHYS-PUB-2019-011}.
This is also useful to measure the properties of the DM multiplet, such as the lifetime and mass spectrum of the components.
Such information is crucial for the determination of the quantum number of the DM and underlying fundamental physics.

%%%%%%%%%%%%%%%%%%%%%%%%%%%%%%%%%%%%%%
\vspace{-.4cm}  
\begin{acknowledgments}
\vspace{-.3cm}
This work is supported in part by JSPS Grant-in-Aid for Scientific Research  No. 17H02878 (S.S.), No. 17K14270 (N.N.), No. 18H05542 (N.N.), No. 18K13535 (S.S.), No. 19H04609 (S.S.) and No. 19K23439 (H.Oide) and by World Premier International Research Center Initiative (WPI Initiative), MEXT, Japan (S.S.) by the Director, Office of Science, Office of
High Energy Physics of the U.S. Department of Energy under the
Contract No. DE-AC02-05CH11231 (H.F.).
\end{acknowledgments}
%%%%%%%%%%%%%%%%%%%%%%%%%%%%%%%%%%%%%%

\bibliography{papers}

\end{document}